\documentclass[aps,pra,reprint,showpacs,superscriptaddress,twocolumn]{revtex4-1}%
\usepackage{amsfonts}
\usepackage{mathrsfs}
\usepackage{amsmath}
\usepackage{amssymb}
\usepackage{graphicx}
\usepackage{color}%
\usepackage[colorlinks,linkcolor=blue,citecolor=blue,hyperindex,bookmarks=false,pdfstartview=FitH]{hyperref}

\providecommand{\U}[1]{\protect\rule{.1in}{.1in}}

\begin{document}
\title{Single-photon nonreciprocal excitation transfer with non-Markovian retarded effects}

\author{Lei Du}
\affiliation{Beijing Computational Science Research Center, Beijing 100193, China}
\author{Mao-Rui Cai}
\affiliation{Beijing Computational Science Research Center, Beijing 100193, China}
\author{Jin-Hui Wu}
\affiliation{Center for Quantum Sciences and School of Physics, Northeast Normal University, Changchun 130024, China}
\author{Zhihai Wang}
\affiliation{Center for Quantum Sciences and School of Physics, Northeast Normal University, Changchun 130024, China}
\author{Yong Li}
\email{liyong@csrc.ac.cn}
\affiliation{Beijing Computational Science Research Center, Beijing 100193, China}
\affiliation{Center for Quantum Sciences and School of Physics, Northeast Normal University, Changchun 130024, China}
\affiliation{Synergetic Innovation Center for Quantum Effects and Applications, Hunan Normal University, Changsha 410081, China}

\date{\today }

%\pacs{42.50.-p, 42.50.Pq, 42.65.-k}

\begin{abstract}
We study at the single-photon level the nonreciprocal excitation transfer between emitters coupled with a common waveguide. Non-Markovian retarded effects are taken into account due to the large separation distance between different emitter-waveguide coupling ports. It is shown that the excitation transfer between the emitters of a small-atom dimer can be obviously nonreciprocal by introducing between them a coherent coupling channel with nontrivial coupling phase. We prove that for dimer models the nonreciprocity cannot coexist with the decoherence-free giant-atom structure although the latter markedly lengthens the lifetime of the emitters. In view of this, we further propose a giant-atom trimer which supports both nonreciprocal transfer (directional circulation) of the excitation and greatly lengthened lifetime. Such a trimer model also exhibits incommensurate emitter-waveguide entanglement for different initial states in which case the excitation transfer is however reciprocal. We believe that the proposals in this paper are of potential applications in large-scale quantum networks and quantum information processing.
\end{abstract}

\maketitle

\section{Introduction}\label{sec1}

Waveguide quantum electrodynamics (QED) studies interactions between atoms and various one-dimensional open waveguides. It provides an excellent platform for achieving strong light-matter interactions due to the strong transverse confinement on the electromagnetic fields~\cite{Wreview1,Wreview2}. Different from cavity QED systems, where atoms are commonly coupled with a single or multiple discrete modes in bounded spaces, atoms can interact with a continuum of modes in waveguides, similar to those of a thermal reservoir~\cite{Wreview2}. In view of this, many disadvantages presented in cavities can be evaded in waveguide QED systems, such as limited bandwidth of emitted photons and stochastic release of cavities~\cite{release}. Since the first experimental realization in 2007~\cite{2007f}, waveguide QED has brought out a great deal of advances, e.g., chiral photon-atom interactions~\cite{chiral1,chiral2,chiral3}, single-photon routers~\cite{rout1,rout2,rout3}, and topologically induced unconventional quantum optics~\cite{tuqo} to name a few. In particular, waveguide-mediated interactions between far apart atoms (resonators), which can be tailored to be either coherent or dissipative, exhibit important applications in achieving large-scale quantum networks~\cite{fan1999,xiao2008,xiao2010,JPan,BBLi,rao}. 

In waveguide QED, atoms are commonly regarded as point-like dipoles because their sizes are in general much smaller than the wavelengths of the waveguide modes they interact with. Recent experiments show that such an approximation is no longer valid when (artificial) atoms interact with a surface acoustic wave whose wavelength can be even much smaller than microwave photons~\cite{transmon1,transmon2,transmon3}. Moreover, it is also possible to couple a single atom with bent waveguides at two or more points separated by distances much larger than one wavelength. Such configurations are referred to as giant atoms~\cite{Lamb1}, which can exhibit striking effects such as frequency-dependent decays and Lamb shifts~\cite{Lamb1,Lamb2}, chiral emission~\cite{chiemit1,chiemit2}, and oscillating bound states~\cite{obs}. Recently, decoherence-free interactions between braided giant atoms are theoretically proposed~\cite{GANori} and experimentally verified~\cite{braided}, where the giant atoms are immune to emitting photons to the waveguide yet they still interact effectively with each other. Moreover, giant-atom structures have also been extended to two or higher dimensions with optical lattices of cold atoms~\cite{highD}. These seminal works provide new inspirations for many applications in quantum simulating and quantum computation.          

It has been shown that non-Markovian retarded effects arising from the large separation distances between, for instance, a single atom and the waveguide end~\cite{end1,end2,end3}, different coupling ports of a giant atom~\cite{longhiretard1}, and far apart atoms~\cite{Solano,longhiretard2,retardresearch} can markedly modify the dynamics. When the traveling time of photons or phonons in the waveguide between different atom-waveguide coupling channels is large enough compared with the inverse of the atomic relaxation rate, the dynamics can exhibit prominent non-Markovianity and thus can not be predicted by common Markovian treatments. This suggests that non-Markovian retarded effect should be taken into account when considering nonlocal couplings that are inevitable in many large-scale systems (e.g., quantum networks). In particular, a counterintuitive phenomenon referred to as ``superradiant paradox'' arises if the separation $L$ between atoms satisfies $l_{c}/2<d<l_{c}$, where $l_{c}$ is the coherent length of photons emitted from the emitters to the waveguide~\cite{Solano,longhiretard2}. As the number of atom increases, the non-Markovianity is shown to be non-negligible even for small separations ~\cite{Natoms}. Such non-Markovian dynamics in single-photon waveguide QED can be solved semi-analytically via a real-space approach~\cite{quantum} and fully analytically via a diagrammatic method~\cite{diag}.  

In this paper, we focus on nonreciprocal excitation transfer between emitters in waveguide QED systems with considerable non-Markovian retarded effects. We start by considering a simple dimer model where two small emitters couple with each other via both direct and indirect (waveguide mediated) interactions. By introducing a nontrivial coupling phase (synthetic magnetic flux) for the direct coupling terms, nonreciprocal excitation transfer can be achieved and the nonreciprocity is also dependent on the phase accumulation of traveling photons in this case. To lengthen the duration of the nonreciprocal phenomenon, we also propose a giant-atom trimer in which the excitation exhibits directional circulation and greatly suppressed dissipation. Dimer models, however, are proved to be incapable of supporting nonreciprocal transfer as long as the emitters become ``decoherence-free''. Moreover, we demonstrate that the entanglement between emitters and the waveguide modes can be quite incommensurate in the case of trivial coupling phase when the single excitation is initially prepared in different emitters, although the excitation transfer is reciprocal in this case.               

\section{Model and equations}\label{sec2}

\begin{figure}[ptb]
\centering
\includegraphics[width=5.5 cm]{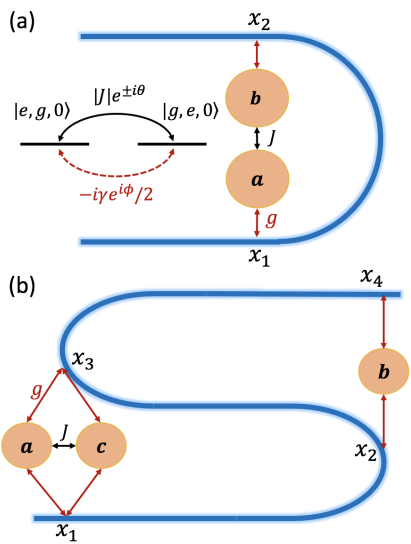}
\caption{Schematic illustration of (a) small-atom dimer and effective energy levels and (b) giant-atom trimer with braided coupling ports.}\label{fig1}
\end{figure}

We first consider a simple model composed of two identical small emitters and a U-type waveguide (the term ``small'' here means that each emitter is coupled with the waveguide at only one point, in contrast to giant-atom models). As shown in Fig.~{\ref{fig1}}, emitters $a$ and $b$ are side-coupled with the waveguide at $x=x_{1}$ and $x=x_{2}$, respectively. In addition, we consider a direct interaction between $a$ and $b$ by assuming that they are spatially close together. Hereafter, we refer to this model as the small-atom dimer for simplicity. The Hamiltonian of the small-atom dimer can be written as ($\hbar=1$)
\begin{equation} 
H=H_{\textrm{e}}+H_{\textrm{w}}+H_{\textrm{int}},
\label{eq1}
\end{equation}
where $H_{\textrm{e}}=\omega_{0}(\sigma_{a}^{+}\sigma_{a}+\sigma_{b}^{+}\sigma_{b})$ and $H_{\textrm{w}}=\sum_{k}\omega_{k}p^{\dag}_{k}p_{k}$ are the free Hamiltonians of the emitters and the waveguide, respectively. Here $\sigma_{j}^{+}$ and $\sigma_{j}$ ($j=a,\,b$) are respectively the raising and lowering operators of emitter $j$ with $\omega_{0}$ the transition frequency between the ground state $|g\rangle$ and excited state $|e\rangle$. $p^{\dag}_{k}$ and $p_{k}$ are respectively the creation and annihilation operators of the traveling photons in the waveguide with wave vector $k$ and frequency $\omega_{k}$. The dispersion relation of the waveguide can be approximately given by $\omega_{k}=v_{g}|k|$ with $v_{g}$ the group velocity if $\omega_{0}$ is far away from the cut-off frequency~\cite{fan2009}. Under the rotating-wave approximation, the interaction Hamiltonian is written as
\begin{eqnarray}
H_{\textrm{int}}=&&\sum_{k}[g_{k}p_{k}(\sigma_{a}^{+}+e^{ikd}\sigma_{b}^{+})+h.c.]\nonumber\\
&&+(J\sigma_{a}^{+}\sigma_{b}+h.c.),
\label{eq2}
\end{eqnarray}  
where $g_{k}$ and $J$ are the emitter-waveguide and emitter-emitter coupling coefficients, respectively, $d=|x_{2}-x_{1}|$ is the separation distance between the two emitters. 

With the Hamiltonian above, the state at time $t$ in the single-excitation manifold can be given by 
\begin{eqnarray}
|\psi(t)\rangle=&&\sum_{k}u_{k}(t)e^{-i\omega_{k}t}p^{\dag}_{k}|g,g,0\rangle\nonumber\\
&&+\sum_{j=a,b}c_{j}(t)\sigma_{j}^{+}e^{-i\omega_{0}t}|g,g,0\rangle,
\label{eq3}
\end{eqnarray}
where $c_{j}(t)$ and $u_{k}(t)$ are the probability amplitudes of exciting emitter $j$ to the excited state and creating a photon with wave vector $k$ in the waveguide, respectively. $|g,g,0\rangle$ denotes the vacuum state of the system with both emitters in the ground state and no photon in the waveguide. By solving the Schr\"{o}dinger equation and eliminating the waveguide modes, one can obtain the time-delayed equations of probability amplitudes (see more details in Appendix~\ref{appa}) 
\begin{equation}
\begin{split}
&\frac{dc_{a}(t)}{dt}=-\frac{\gamma}{2}e^{i\phi}c_{b}(t-\frac{d}{v_{g}})\Theta(t-\frac{d}{v_{g}})\\
&\qquad\qquad-\frac{\kappa+\gamma}{2}c_{a}(t)-iJc_{b}(t),\\
&\frac{dc_{b}(t)}{dt}=-\frac{\gamma}{2}e^{i\phi}c_{a}(t-\frac{d}{v_{g}})\Theta(t-\frac{d}{v_{g}})\\
&\qquad\qquad-\frac{\kappa+\gamma}{2}c_{b}(t)-iJ^{*}c_{a}(t),
\end{split}
\label{eq4}
\end{equation}
where $\gamma=2|g_{k_{0}}|^{2}/v_{g}$ is the spontaneous emission rate of the emitters to the waveguide with $k_{0}=\omega_{0}/v_{g}$~\cite{Solano}. $\phi=k_{0}d$ is the phase accumulation of photons traveling from one emitter to another through the waveguide and $\Theta(t)$ is the Heaviside step function. $\kappa$ denotes the loss of the emitters due to other decay channels, which can be much smaller than $\gamma$ experimentally. 

Equation~(\ref{eq4}) shows that the waveguide introduces both a decay channel for each emitter and a retarded indirect coupling between them. In the Markovian limit $d/v_{g}\rightarrow0$ and for long-time evolution $t\rightarrow\infty$, Eq.~(\ref{eq4}) can be approximately regarded as simultaneous differential equations, while as $d$ increases gradually, the non-Markovian retarded effect becomes more and more dominant such that the dynamic evolution can markedly deviate from the Markovian expectation. 

Note that the coupling phase $\theta$ of the direct interaction should be considered (i.e., $J=|J|e^{i\theta}$), which can not be removed by any gauge transformation in the presence of the waveguide mediated coupling. As will be shown in the following, it plays a crucial role for achieving nonreciprocal excitation transfer. Experimentally, such a coupling phase can be achieved via an ac driving in each emitter~\cite{KFang,Roushan,acdrive1,acdrive2}. For whispering gallery mode resonators, one can also use an anti-resonant linker to introduce an optical path imbalance $\Delta x$ in opposite directions between the two resonators, such that the effective coupling phase reads $\theta=2\pi\Delta x/\lambda$ with $\lambda$ the resonant wavelength of the resonators~\cite{hafezi1,hafezi2,ljin1,ljin2}.

\section{Nonreciprocal excitation transfer}\label{sec3}
\begin{figure}[ptb]
\centering
\includegraphics[width=8 cm]{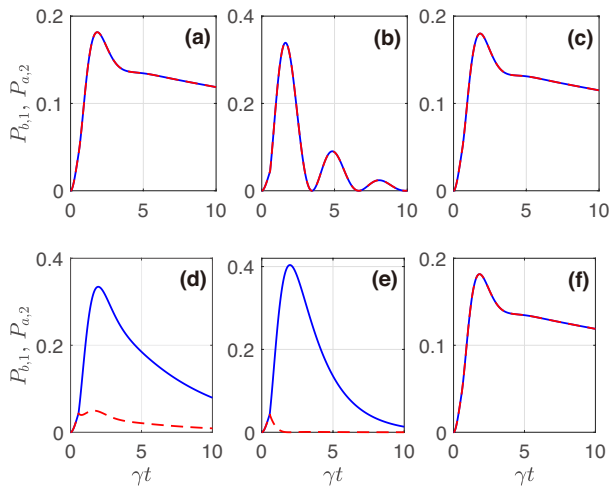}
\caption{Dynamic evolutions of populations $P_{b,1}$ (blue solid) and $P_{a,2}$ (red dashed) for (a) $\eta=0.56$, (b) $\eta=0.574$, (c) $\eta=0.588$, (d) $\theta=\pi/4$, (e) $\theta=\pi/2$, (f) $\theta=\pi$. Here we assume $\theta=0$ in (a)-(c) and $\eta=0.56$ in (d)-(f). Other parameters are $\omega_{0}/\gamma=112.19$, $\kappa/\gamma=8.7\times10^{-3}$, and $|J|/\gamma=0.5$.}\label{fig2}
\end{figure}

Now we consider two initial states $|\psi_{1}(0)\rangle=|e,g,0\rangle$ and $|\psi_{2}(0)\rangle=|g,e,0\rangle$ (either $a$ or $b$ is initially prepared in the excited state) to compare the excitation transfers from $a$ to $b$ and from $b$ to $a$, respectively. This can be done by focusing on the dynamic evolutions of the probabilities $P_{b,1}$ and $P_{a,2}$, where $P_{j,n}$ ($j=a,\,b$; $n=1,\,2$) denotes the populations $|c_{j}|^{2}$ of emitter $j$ with initial state $|\psi_{n}(0)\rangle$. Moreover, we define $\eta=d\gamma/v_{g}$ as the separation distance between the emitters normalized by the coherence length~\cite{longhiretard1,Solano,longhiretard2}. Therefore the relation between phase $\phi$ and the time delay reads $\phi=\omega_{0}d/v_{g}=\omega_{0}\eta/\gamma$. For example, if emitters $a$ and $b$ are two identical superconducting qubits (artificial atoms) with $\omega_{0}/2\pi=3.276\textrm{GHz}$ and $\gamma/2\pi=29.2\textrm{MHz}$, we have $\phi=\{20,\,20.5,\,21\}\pi$ for $\eta=\{0.56,\,0.574,\,0.588\}$, respectively.  

We first consider the case of trivial coupling phase $\theta=0$ and plot in Figs.~\ref{fig2}(a)-\ref{fig2}(c) the dynamical evolutions of $P_{b,1}$ and $P_{a,2}$ with $\eta=0.56$, $\eta=0.574$, and $\eta=0.588$, respectively. It shows that the waveguide induced phase factor $\phi$ cannot lead to nonreciprocity since it does not break the time-reversal symmetry of the Hamiltonian. Note that the populations decay much slower when $\phi$ is an integer multiple of $\pi$. This is reminiscent of the Fabry-P\'erot bound states in the continuum (BICs) in the Markovian limit, which shows that one of the eigenstates becomes lossless if $\phi=m\pi$ ($m$ is an arbitrary integer)~\cite{BIC1,BIC2}. For general initial states considered here, one can find from Eq.~(\ref{eq4}) that the waveguide induced indirect coupling is purely dissipative (i.e., $i\gamma e^{i\phi}/2$ is purely imaginary) in the case of $\phi=m\pi$, which serves as an effective gain and thus suppresses the decay of the emitters~\cite{UAPT}. Moreover, the populations decay in an oscillating form for $\phi\neq m\pi$ because the excitation bounces between the emitters back and forth. 

On the other hand, Figs.~\ref{fig2}(d) and \ref{fig2}(e) exhibit obvious nonreciprocal excitation transfer within a certain time range due to the nontrivial coupling phase ($\theta\neq m\pi$), which breaks the time-reversal symmetry of the Hamiltonian. In particular, the optimal nonreciprocal transfer can be achieved for $\theta=\pi/2$ (the maximum of $|P_{b,1}(t)-P_{a,2}(t)|$ during the evolution maximizes for $\theta=\pi/2$). As shown in the effective energy-level diagram in Fig.~\ref{fig1}(a), the nontrivial coupling phase $\theta$ has no impact on dynamics for $t<d/v_{g}$ because it can always be gauged away in the absence of the retarded coupling (denoted by the red dashed line). This is also why $P_{b,1}$ and $P_{a,2}$ coincide with each other in the beginning. Once each emitter meets the retarded feedback coming from the other one at $t=d/v_{g}$, an additional transfer path between two emitters is formed such that the two paths can interfere with each other (see the black solid and red dashed lines in the energy-level diagram) and the interference effects of opposite directions are generally different for $\theta\neq m\pi$. Figure~\ref{fig2}(f) shows that the transfer becomes reciprocal again for $\theta=\pi$, which attributes to the recovered time-reversal symmetry. 

\begin{figure}[ptb]
\centering
\includegraphics[width=8 cm]{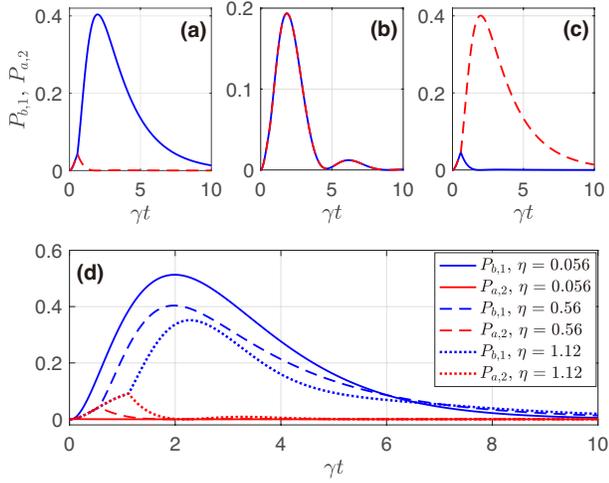}
\caption{(a)-(c): Dynamic evolutions of populations $P_{b,1}$ (blue solid) and $P_{a,2}$ (red dashed) for (a) $\eta=0.56$, (b) $\eta=0.574$, and (c) $\eta=0.588$. (d) Dynamic evolutions of populations $P_{b,1}$ and $P_{a,2}$ with different values of $\eta$ corresponding to $\phi=2m\pi$. Here we assume $\theta=\pi/2$ in all panels and other parameters are the same as those in Fig.~\ref{fig2}.}\label{fig3}
\end{figure}

Although the time-reversal symmetry of the system is broken by tuning the coupling phase $\theta$, the nonreciprocity is also dependent on $\phi$ in the case of $\theta\neq m\pi$. As shown in Figs.~\ref{fig3}(a)-(c), one can observe nearly inverse nonreciprocal transfer for $\phi=20\pi$ and $\phi=21\pi$ ($\eta=0.56$ and $\eta=0.588$), while the transfer becomes reciprocal for $\phi=20.5\pi$ ($\eta=0.574$) even if the coupling phase is nontrivial. This is because the coupling phases are effectively shifted from $\pm\theta$ to $\pm\theta-\phi$ by removing $\phi$ in the indirect coupling terms, implying that the moduli of the overall couplings of opposite directions are effectively swapped in the case of $\phi=(2m+1)\pi$. Note that the nonreciprocal behaviors are not exactly inverse for $\phi=20\pi$ and $21\pi$ due to different time delays before which the emitters decay exponentially~\cite{Solano}.

We point out that $\eta$ determines the non-Markovianity of the system such that it also affects the onset time and the optimal effect of the nonreciprocal transfer. As shown in Fig. ~\ref{fig3}(d), the onset time of the nonreciprocal transfer is exactly $t=d/v_{g}$ (i.e., $\gamma t=\eta$). Moreover, all values of $\eta$ chosen here correspond to $\phi=2m\pi$ ($\phi=\{2\pi,\,20\pi,\,40\pi\}$ for $\eta=\{0.056,\,0.56,\,1.12\}$), which yield the optimal nonreciprocal transfer for $\theta=\pi/2$ as discussed above. It shows that the optimal effect of the nonreciprocal transfer becomes worse and worse as $\eta$ increases. In other words, the retarded effect puts off the onset and suppresses the degree of the nonreciprocal excitation transfer. 

Note that in the absence of external inputs, the populations of both emitters should fall to zero rapidly and the nonreciprocal phenomenon can only be observed within a short-lived duration, as shown in Figs.~\ref{fig2} and \ref{fig3}. It has been shown that giant atoms (self-interference resonators) can be completely decoupled from the waveguide and thus no longer emit photons to it~\cite{GANori,braided,leidu}. However, this generally makes the emitters isolated such that they can hardly interact with each other if they are spatially separated. Thanks to the braided structure proposed in Refs.~\cite{GANori,braided}, decoherence-free couplings can be achieved between far apart giant atoms, i.e., the spontaneous emissions of the atoms to the waveguide can be completely suppressed while the indirect coupling between them is nonvanishing. Nevertheless, we would like to point out that a dimer model with such a braided structure is unable to demonstrate nonreciprocal transfer, although the lifetime of the emitter can be markedly extended in this case (see more details in Appendix~\ref{appb}).       

\section{Directional excitation circulation in a giant-atom trimer}\label{sec3}
As discussed in Sec.~\ref{sec2} and Appendix~\ref{appb}, nonreciprocal excitation transfer is not allowed in dimer models with ``decoherence-free'' giant atoms (the quotation mark here means that the giant atoms are not exactly decoherence-free due to the retarded self-interference effects), although the lifetime of the emitters can be markedly lengthened. In view of this, we extend the braided structure by introducing the third emitter $c$ (with the raising and lowering operators denoted by $\sigma_{c}^{+}$ and $\sigma_{c}$, respectively). As shown in Fig.~\ref{fig1}(b), emitters $a$ and $c$ are coupled with the waveguide via the same two ports located at $x=x_{1}$ and $x=x_{3}$, respectively, while emitter $b$ couples with the waveguide at $x=x_{2}$ and $x=x_{4}$, respectively. The four coupling ports are arranged in the braided manner to suppress the spontaneous emission to the waveguide and obtain nonvanishing indirect coupling. We assume that the coupling ports are evenly spaced (i.e., $x_{2}-x_{1}=x_{3}-x_{2}=x_{4}-x_{3}=d$) and all emitter-waveguide couplings are identical. Moreover, $a$ and $c$ couple directly with each other in terms of $|J|e^{i\theta}\sigma_{a}^{+}\sigma_{c}+h.c.$. For simplicity, we refer to this structure as the giant-atom trimer hereafter. In this case, the effective equations of the probability amplitudes are written as  
\begin{equation}
\begin{split}
\frac{dc_{a}(t)}{dt}=&-\frac{\gamma}{2}(3D_{1,b}+D_{3,b})-\gamma(D_{2,a}+D_{2,c})\\
&-(\kappa+\gamma)a(t)-i(|J|e^{i\theta}-i\gamma)c_{c}(t),\\
\frac{dc_{b}(t)}{dt}=&-\frac{\gamma}{2}[3(D_{1,a}+D_{1,c})+D_{3,a}+D_{3,c}]\\
&-(\kappa+\gamma)b(t)-\gamma D_{2,b},\\
\frac{dc_{c}(t)}{dt}=&-\frac{\gamma}{2}(3D_{1,b}+D_{3,b})-\gamma(D_{2,a}+D_{2,c})\\
&-(\kappa+\gamma)c(t)-i(|J|e^{-i\theta}-i\gamma)c_{a}(t),
\end{split}
\label{eq5}
\end{equation}  
where $D_{n,l}=c_{l}(t-nd/v_{g})e^{in\phi}\Theta(t-nd/v_{g})$ ($n=1,\,2,\,3$; $l=a,\,b,\,c$) with subscript $n$ corresponding to time delay $nd/v_{g}$. Equation~(\ref{eq5}) shows that there are three different coupling channels between emitters $a$ and $c$, including two indirect couplings with one of them being retarded and one direct coupling. Clearly, the overall coupling is asymmetric as long as $\theta\neq m\pi$ and the asymmetry is maximized when $J=\gamma$ and $\theta=(m+1/2)\pi$ (for this reason, we choose $J=\gamma$ in this case).  

\begin{figure}[ptb]
\centering
\includegraphics[width=8 cm]{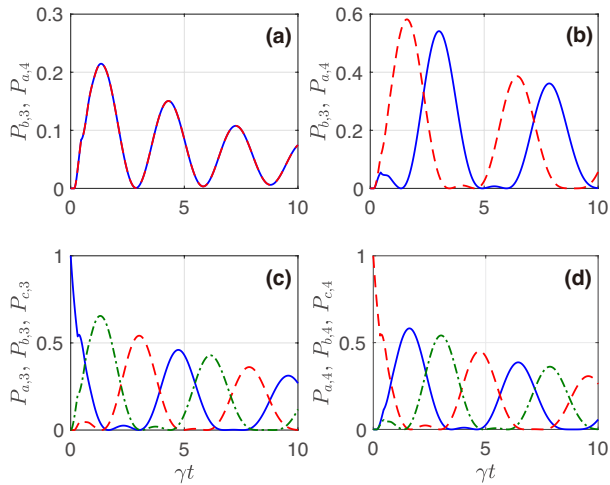}
\caption{(a) and (b): Dynamic evolutions of populations $P_{b,3}$ (blue solid) and $P_{a,4}$ (red dashed) for (a) $\theta=0$ and (b) $\theta=\pi/2$. (c) and (d): Dynamic evolutions of populations of all three emitters with (c) $\theta=\pi/2$ and initial state $|\psi_{3}(0)\rangle$, (d) $\theta=\pi/2$ and initial state $|\psi_{4}(0)\rangle$. Other parameters are the same as those in Fig.~\ref{fig3}(a) except for $|J|/\gamma=1$ ($|J|$ and $\theta$ denote in this case the amplitude and phase of the direct coupling coefficient between $a$ and $c$, respectively).}\label{fig4}
\end{figure}

We plot in Figs.~\ref{fig4}(a) and \ref{fig4}(b) the dynamic evolutions of the populations $P_{b,3}$ and $P_{a,4}$, respectively, where $P_{j,3(4)}$ ($j=a,\,b$) denotes the population $|c_{j}|^{2}$ of emitter $j$ with the initial state $|\psi_{3}(0)\rangle=|e,g,g,0\rangle$ ($|\psi_{4}(0)\rangle=|g,e,g,0\rangle$). We find that in the giant-atom trimer, nonreciprocal excitation transfer emerges again for $\theta\neq m\pi$ due to the asymmetric overall coupling between $a$ and $c$. Different from the behaviors in Figs.~\ref{fig2} and \ref{fig3}, $P_{b,3}$ and $P_{a,4}$ oscillate here with a fixed phase difference. To understand this difference, we also plot in Figs.~\ref{fig4}(b) and \ref{fig4}(c) the evolutions of the populations of all three emitters in the case of $\theta=\pi/2$, with initial states $|\psi_{3}(0)\rangle$ and $|\psi_{4}(0)\rangle$, respectively. One can find that for both initial states, the excitation exhibits a directional circulation along the same direction of $a\rightarrow c\rightarrow b\rightarrow a$ after $t=d/v_{g}$, which is a signature of broken time-reversal symmetry that cannot be observed for $\theta=m\pi$. That is to say, the single excitation initially prepared in emitter $b$ is preferentially transferred to $a$ while that initially prepared in emitter $a$ shows a preferential transfer to $c$. Note that circulations along the opposite direction $a\rightarrow b\rightarrow c\rightarrow a$ can be achieved for $\theta=-\pi/2$. This is analogous to chiral currents of electrons in an Aharonov-Bohm cage and of photons in a synthetic magnetic field~\cite{chiralcurrent,sgspin,XXW1,XXW2}. In this way, the target emitter ($b$ for $|\psi_{3}(0)\rangle$ and $a$ for $|\psi_{4}(0)\rangle$) can be excited efficiently within different durations for the two initial states. Note that another initial state $|\psi_{5}(0)\rangle=|g,g,e,0\rangle$ results in essentially the same dynamics as those with $|\psi_{3}(0)\rangle$. This is because Eq.~(\ref{eq5}) is invariant by exchanging the amplitudes $c_{a}$ and $c_{c}$ and reversing the coupling phase $\theta$ (with identical $\theta$, $|\psi_{3}(0)\rangle$ and $|\psi_{5}(0)\rangle$ lead to exactly reverse circulations). Moreover, since emitter $b$ is not requested to interact directly with $a$ or $c$ in this case, the trimer model can be used for nonreciprocal excitation transfer between far apart emitters, which is of potential applications in large-scale quantum networks.

\begin{figure}[ptb]
\centering
\includegraphics[width=8.5 cm]{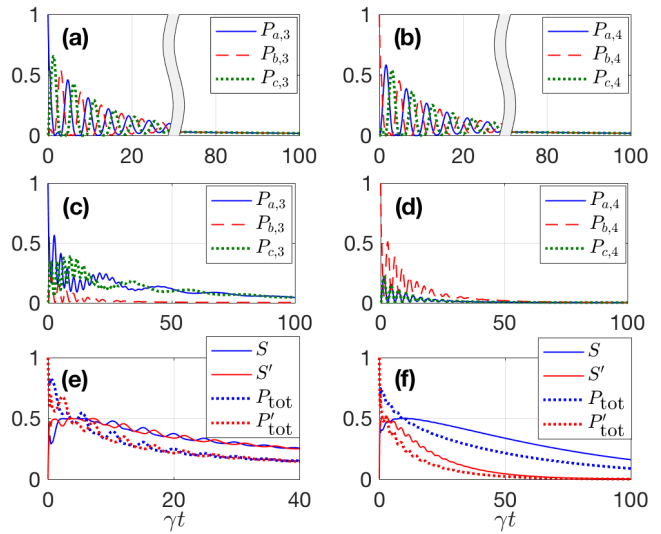}
\caption{(a)-(d): Dynamic evolutions of populations of all three emitters with (a) $\theta=\pi/2$ and initial state $|\psi_{3}(0)\rangle$, (b) $\theta=\pi/2$ and initial state $|\psi_{4}(0)\rangle$, (c) $\theta=0$ and initial state $|\psi_{3}(0)\rangle$, (d) $\theta=0$ and initial state $|\psi_{4}(0)\rangle$. (e) and (f): Dynamic evolutions of linear entropies and total populations with (e) $\theta=\pi/2$ and (f) $\theta=0$. Other parameters are the same as those in Fig.~\ref{fig4}.}\label{fig5}
\end{figure}

Finally, we focus on the dynamic evolutions of the giant-atom trimer in long-time limit. We first plot in Figs.~\ref{fig5}(a)-\ref{fig5}(d) the evolutions of the populations of all three emitters with different values of $\theta$ and initial states. Two major differences between the cases of $\theta=\pi/2$ and $\theta=0$ can be found: (i) for $\theta=\pi/2$ the populations of all three emitters decay in similar oscillating manners and tend to vanish together, while for $\theta=0$ the population of $b$ exhibits obviously different evolution from those of $a$ and $c$; (ii) for $\theta=\pi/2$ the long-time evolutions of the populations are similar for both initial states [see Figs.~\ref{fig5}(a) and \ref{fig5}(b)] while it is not true for $\theta=0$, i.e., the long-time evolutions are quite different for initial states $|\psi_{3}(0)\rangle$ and $|\psi_{4}(0)\rangle$ [see Figs.~\ref{fig5}(c) and \ref{fig5}(d)]. Physically, this is because for $\theta=\pi/2$, the giant-atom trimer exhibits directional excitation circulation as mentioned above, yet the oscillation amplitudes of all three populations minish gradually due to both the other decay channels and the non-Markovian retarded effect (as discussed in Appendix~\ref{appb}, the decoherence of the emitters to the waveguide can be exactly suppressed in case of $\kappa\rightarrow0$ and $d/v_{g}\rightarrow0$ such that the populations oscillate stably). For $\theta=0$, the excitation can be transferred simultaneously from one emitter to both the other two. The transfer probability from $b$ to $a$ and that from $b$ to $c$ should be equal because the two paths are identical. However, the transfer probability from $a$ ($c$) to $b$ and that from $a$ ($c$) to $c$ ($a$) are unequal because the two paths are quite different [the overall coupling between $a$ ($c$) and $b$ is purely coherent with time delays $d/v_{g}$ and $3d/v_{g}$ while that of $a$ and $c$ contains both coherent and dissipative parts with time delays $0$ or $2d/v_{g}$]. For initial state $|\psi_{3}(0)\rangle$, the excitation is apt to be transferred from $a$ to $c$ rather than from $a$ to $b$ while it comes back equiprobably from $b$ to $a$ and from $b$ to $c$. As a result, the excitation tends to bounce between $a$ and $c$ in long-time limit. For initial state $|\psi_{4}(0)\rangle$, the populations of $a$ and $c$ are always identical because the excitation initialized in $b$ is transferred between $b$ and $a$ or between $b$ and $c$ with identical probabilities since the beginning.         

We point out that the entanglement between the emitters and the waveguide mode can be markedly incommensurate for initial states $|\psi_{3}(0)\rangle$ and $|\psi_{4}(0)\rangle$ in the case of $\theta=0$. This can be verified by calculating the linear entropy $S=1-\textrm{Tr}(\rho^{2})$, which estimates here the correlation between the emitters and the electromagnetic field in the waveguide~\cite{Solano,maxedness}. $\rho$ denotes the reduced density matrix of the emitters, which can be obtained by taking a trace over the waveguide states, i.e., $\rho=\textrm{Tr}_{\textrm{w}}[|\psi(t)\rangle\langle\psi(t)|]$ with $|\psi(t)\rangle$ given in Eq.~(\ref{eq3}). We plot in Figs.~\ref{fig5}(e) and \ref{fig5}(f) the linear entropies $S$ (corresponding to initial state $|\psi_{3}(0)\rangle$) and $S'$ (corresponding to initial state $|\psi_{4}(0)\rangle$) for $\theta=\pi/2$ and $\theta=0$, respectively. It can be found that for $\theta=0$ the evolutions of linear entropy can be quite different by initially exciting different emitters ($a$ or $b$) to the excited state while for $\theta=\pi/2$ there is no significant difference between the two cases. We also plot in Figs.~\ref{fig5}(e) and \ref{fig5}(f) the total populations of all emitters (i.e., survival probabilities) $P_{\textrm{tot}}=P_{a,3}+P_{b,3}+P_{c,3}$ and $P_{\textrm{tot}}'=P_{a,4}+P_{b,4}+P_{c,4}$ for $\theta=\pi/2$ and $\theta=0$, respectively. As expected, the total populations exhibit similar behaviors with those of the linear entropies~\cite{Solano}. Such a result is reminiscent of the nonreciprocal entanglement demonstrated in Ref.~\cite{neJH}, which reveals that the conditions of nonreciprocal entanglement and transport are not necessary the same. However, we refer to the phenomenon here as incommensurate entanglement to distinguish it from the phenomena stemming from broken time-reversal symmetry. As a final note, the incommensurate entanglement vanishes in the Markovian limit, where for $\theta=m\pi$ the excitation is transferred equally from the initial emitter to the other two and comes back with the same probability, independent of the initial state.    

\section{Conclusions}
In summary, we have studied the nonreciprocal excitation transfer in the presence of nonnegligible non-Markovian retarded effects by considering two waveguide QED models, i.e., a small-atom dimer and a giant-atom trimer. Both models exhibit nonreciprocal single-excitation transfer if a coherent coupling channel with nontrivial coupling phase is introduced, while the waveguide induced phase accumulation does not break the time-reversal symmetry and thereby cannot result in nonreciprocity solely. The retarded effects, which are inevitable in the presence of nonlocal couplings, are shown to put off the onset and suppress the degree of the nonreciprocal transfer. In particular, we have demonstrated that the giant-atom trimer supports both nonreciprocal excitation transfer (in a directionally circulatory manner) and greatly suppressed decoherence of the emitters, which cannot be achieved in dimer models with similar decoherence-free structures. Moreover, incommensurate emitter-waveguide entanglement has been revealed when different emitters of the giant-atom trimer are initially excited, whose condition is independent of the time-reversal symmetry. The results in this paper may inspire applications based on large-scale quantum networks due to the rapid progress in relevant experimental platforms.   

\section*{Acknowledgments}

L. Du thanks D.-W. Xiao and Q.-S. Zhang for helpful discussions. This work was supported by the Science Challenge Project (Grant No. TZ2018003), the National Key R\&D Program of China (Grant No. 2016YFA0301200), and the National Natural Science Foundation of China (Grants No. 11774024, No. 11875011, No. 12047566, No. 12074030, No. 12074061, and No. U1930402).   

\appendix
\section{Derivation of Eq.~(\ref{eq4})}\label{appa}
In this appendix, we show in detail the derivation of Eq.~(\ref{eq4}) in the main text. With the Hamiltonians and the single-excitation wave function given in Eqs.~(\ref{eq1})-(\ref{eq3}) and solving the schr\"{o}dinger equation, one can obtain the equations of the probability amplitudes
\begin{equation}
\begin{split}
&\frac{dc_{a}(t)}{dt}=-i\sum_{k} g_{k}u_{k}(t)e^{-i(\omega_{k}-\omega_{0})t}\\
&\qquad\qquad-iJc_{b}(t),\\
&\frac{dc_{b}(t)}{dt}=-i\sum_{k} g_{k}u_{k}(t)e^{ikd}e^{-i(\omega_{k}-\omega_{0})t}\\
&\qquad\qquad-iJ^{*}c_{a}(t),\\
&\frac{du_{k}(t)}{dt}=-ig^{*}_{k}e^{i(\omega_{k}-\omega_{0})t}[c_{a}(t)+e^{-ikd}c_{b}(t)].
\end{split}
\label{eqa1}
\end{equation}
The formal solution of $u_{k}(t)$ can be written as
\begin{eqnarray}
u_{k}(t)=&&-i\int_{t_{0}}^{t}dt' g^{*}_{k}e^{i(\omega_{k}-\omega_{0})t'}[c_{a}(t')\nonumber\\
&&+e^{-ikd}c_{b}(t')],
\label{eqa2}
\end{eqnarray}
where $t_{0}<t$ is the initial time and $u_{k}(t_{0})=0$ represents the initial vacuum state of the waveguide. Substituting Eq.~(\ref{eqa2}) into Eq.~(\ref{eqa1}), one has 
\begin{equation}
\begin{split}
\frac{dc_{a}(t)}{dt}=&-iJc_{b}(t)-\frac{1}{2\pi}\int_{t_{0}}^{t}dt' e^{i\omega_{0}(t-t')}\int_{-\infty}^{+\infty}dk\\
&\times|g_{k}|^{2}[c_{a}(t')+e^{-ikd}c_{b}(t')]e^{-i\omega_{k}(t-t')},\\
\frac{dc_{b}(t)}{dt}=&-iJ^{*}c_{a}(t)-\frac{1}{2\pi}\int_{t_{0}}^{t}dt' e^{i\omega_{0}(t-t')}\int_{-\infty}^{+\infty}dk\\
&\times|g_{k}|^{2}[e^{ikd}c_{a}(t')+c_{b}(t')]e^{-i\omega_{k}(t-t')}.
\end{split}
\label{eqa3}
\end{equation} 
Considering that both $\omega_{k}$ and $g_{k}$ are even functions of $k$, i.e., $\omega(k)=\omega(-k)$ and $g(k)=g(-k)$, one can change the variable that is being integrated, i.e.,
\begin{equation}
\begin{split}
\frac{dc_{a}(t)}{dt}=&-iJc_{b}(t)-\int_{t_{0}}^{t}dt' e^{i\omega_{0}(t-t')}\int_{0}^{+\infty}d\omega \frac{|g_{k}|^{2}}{2\pi v_{g}}\\
&\times[2c_{a}(t')+2\cos{(kd)}c_{b}(t')]e^{-i\omega(t-t')},\\
\frac{dc_{b}(t)}{dt}=&-iJ^{*}c_{a}(t)-\int_{t_{0}}^{t}dt' e^{i\omega_{0}(t-t')}\int_{0}^{+\infty}d\omega \frac{|g_{k}|^{2}}{2\pi v_{g}}\\
&\times[2\cos{(kd)}c_{a}(t')+2c_{b}(t')]e^{-i\omega(t-t')},
\end{split}
\label{eqa4}
\end{equation} 
where we have denoted $\omega_{k}$ by $\omega$ for simplicity. Note that one has $|g_{k}|^{2}/v_{g}=|g_{k}|^{2}(dk/d\omega)=2\pi|g_{\omega}|^{2}D_{\omega}$ according to the Fermi's golden rule~\cite{Fermi,PSolano}, where $g_{\omega}$ is the emitter-waveguide coupling coefficient as a function of $\omega$ and $D_{\omega}$ is the density of states of the frequency space~\cite{Lamb1,braided}. Assuming that $|g(\omega)|^{2}/v_{g}=\gamma/2$ is constant under the Markovian approximation~\cite{longhiretard1,Solano,longhiretard2} and $\omega_{0}$ is far away from the cut-off frequency of the waveguide such that its dispersion can be approximately linearized as $\omega=\omega_{0}+\nu$ if ~\cite{fan2009,wangnr}, we have
\begin{equation}
\begin{split}
&\frac{dc_{a}(t)}{dt}=-\frac{\gamma}{2\pi}\int_{-\infty}^{+\infty}d\nu\int_{t_{0}}^{t}dt'e^{-i\nu(t-t')}\{c_{a}(t')\\
&\qquad\qquad+\cos{[(k_{0}+\frac{\nu}{v_{g}})d]}c_{b}(t')\}-iJc_{b}(t),\\
&\frac{dc_{b}(t)}{dt}=-\frac{\gamma}{2\pi}\int_{-\infty}^{+\infty}d\nu\int_{t_{0}}^{t}dt'e^{-i\nu(t-t')}\{c_{b}(t')\\
&\qquad\qquad+\cos{[(k_{0}+\frac{\nu}{v_{g}})d]}c_{a}(t')\}-iJ^{*}c_{a}(t),
\end{split}
\label{eqa5}
\end{equation}
where $k_{0}=\omega_{0}/v_{g}$. According to the definition of the delta function $\int d\omega e^{i\omega t}=2\pi\delta(t)$, Eq.~(\ref{eqa5}) can be simplified as Eq.~(\ref{eq4}) in the main text by including the decay $\kappa$ for each emitter. Note that we have dropped the contribution of $\delta(t'-t-d/v_{g})$ at this step because it is centered outside the the range of integral, i.e., $t'-t-d/v_{g}$ is negative for $t'\in[0,t]$. This is always true as long as $d$ is not exactly zero, even if $d$ is very small. 

\section{Extended lifetime with braided giant-atom structure}\label{appb}

\begin{figure}[ptb]
\renewcommand {\thefigure} {B1}
\centering
\includegraphics[width=5.5 cm]{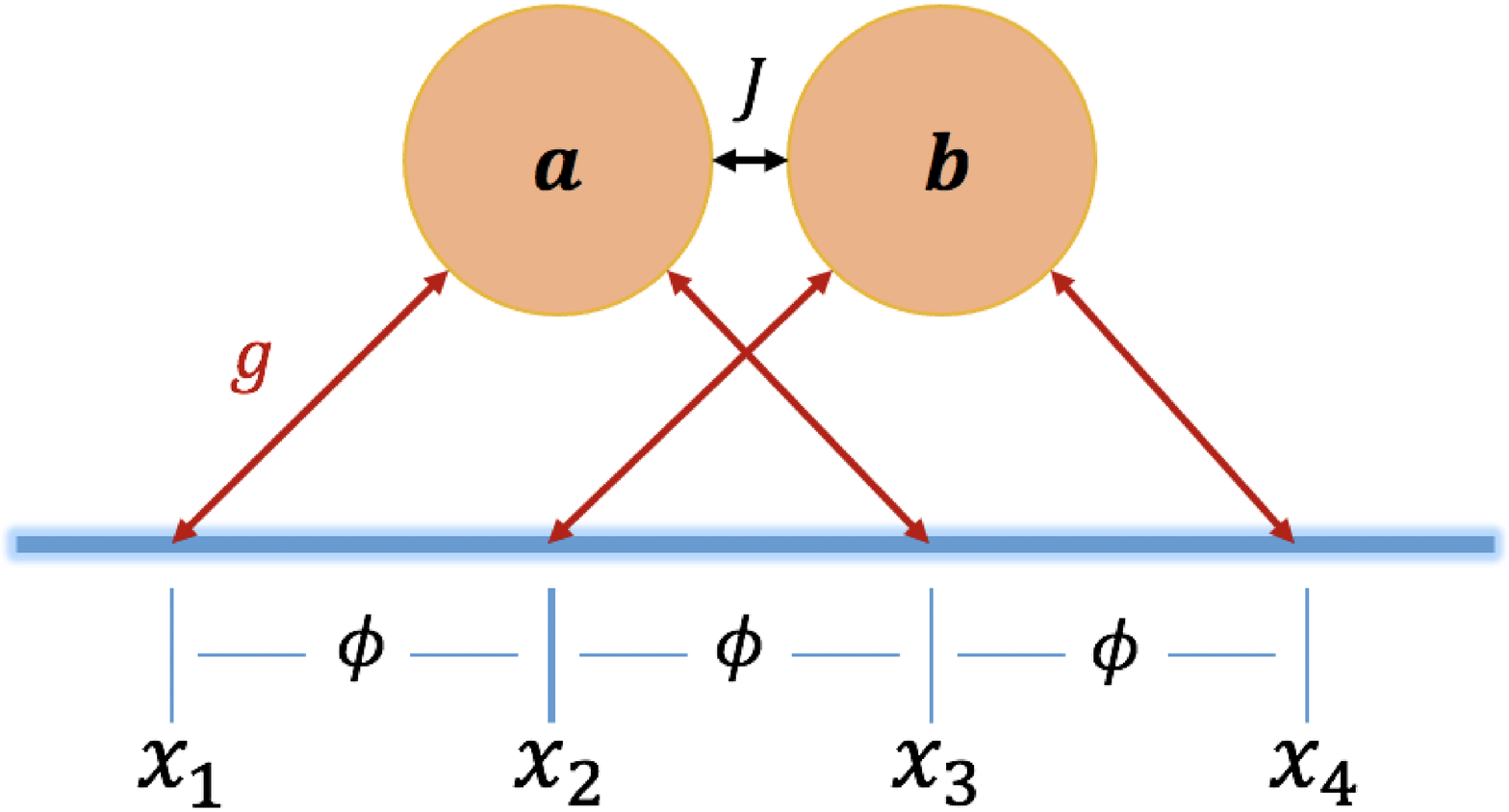}
\caption{Schematic illustration of braided giant-atom dimer.}\label{figb}
\end{figure}

In this appendix, we aim to prove that two giant atoms with decoherence-free indirect coupling cannot exhibit nonreciprocal transfer even in the presence of a direct coupling between them. As shown in Fig.~\ref{figb}, both emitters $a$ and $b$ couple with the waveguide twice in the braided manner, i.e., emitter $a$ ($b$) couples with the waveguide at $x=x_{1}$ and $x=x_{3}$ ($x=x_{2}$ and $x=x_{4}$). The emitter-waveguide couplings are assumed to be identical (i.e., $g$) and the coupling ports are evenly spaced by distance $d$. Such a braided structure can be implemented, for example, with a S-type waveguide~\cite{GANori,braided}. To distinguish from the small-atom dimer in Fig.~\ref{fig1}(a), we refer to this model as the giant-atom dimer. Once again, we consider the direct coupling $|J|e^{\pm i\theta}$ between $a$ and $b$. In this case, the effective equations of the probability amplitudes $c_{a}$ and $c_{b}$ are written as       
\begin{equation}
\begin{split}
&\frac{dc_{a}(t)}{dt}=-\gamma c_{a}(t)-\frac{\gamma}{2}(3D_{1,b}+D_{3,b})\\
&\qquad\qquad-\gamma D_{2,a}-i|J|e^{i\theta}c_{b}(t),\\
&\frac{dc_{b}(t)}{dt}=-\gamma c_{b}(t)-\frac{\gamma}{2}(3D_{1,a}+D_{3,a})\\
&\qquad\qquad-\gamma D_{2,b}-i|J|e^{-i\theta}c_{a}(t),
\end{split}
\label{eqb1}
\end{equation}
where $\phi=k_{0}d=\omega_{0}\eta$ and $D_{n,l}=c_{l}(t-nd/v_{g})e^{in\phi}\Theta(t-nd/v_{g})$ ($n=1,\,2,\,3$; $l=a,\,b$). Here we have neglected other decay channels for simplicity which can be experimentally much weaker than the spontaneous emission to the waveguide. 

\begin{figure}[ptb]
\renewcommand {\thefigure} {B2}
\centering
\includegraphics[width=8 cm]{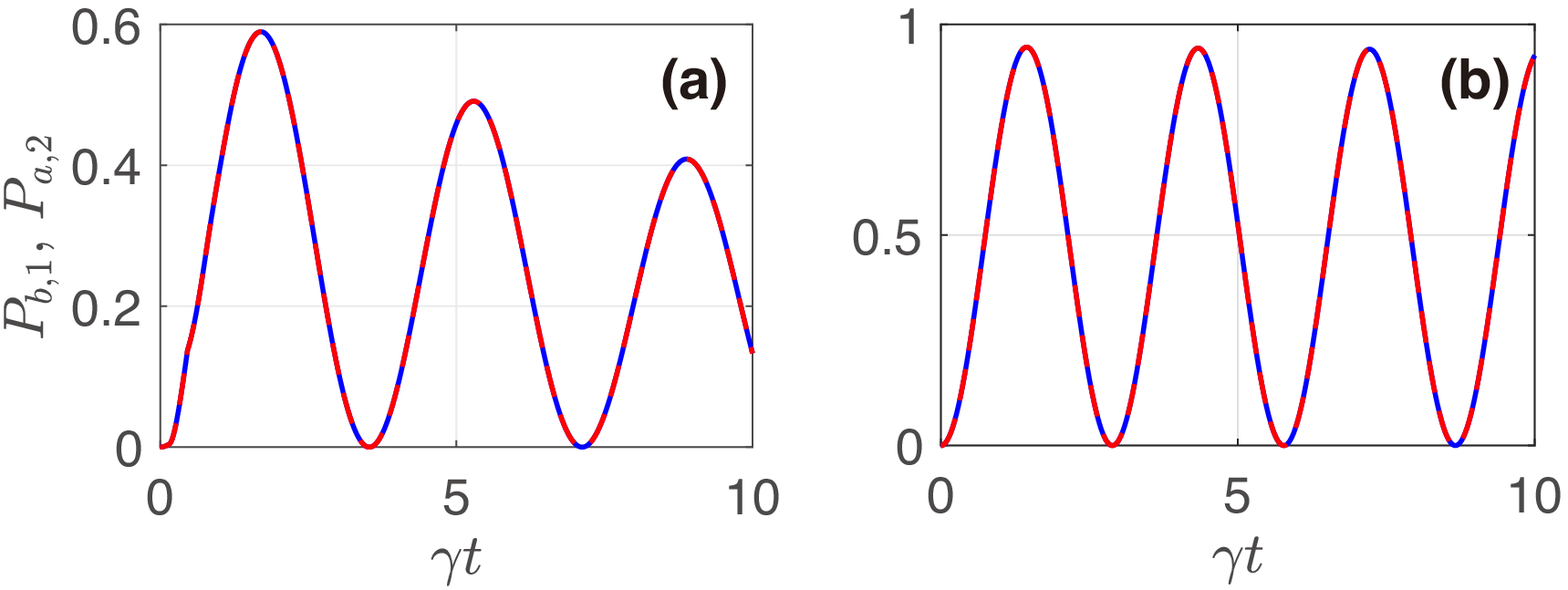}
\caption{Dynamic evolution of populations $P_{b,1}$ (blue solid) and $P_{a,2}$ (red dashed) with (a) $\eta=0.154$ ($\phi=5.5\pi$) and (b) $\eta=0.014$ ($\phi=0.5\pi$). Other parameters are the same as those in Fig.~\ref{fig3}.}\label{figb2}
\end{figure}

We plot in Fig.~\ref{figb2}(a) the dynamic evolutions of $P_{b,1}$ and $P_{a,2}$ by numerically calculating Eq.~(\ref{eqb1}). The separation distance $d$ between adjacent coupling ports is well tailored such that $2\phi=\pi$. According to Ref.~\cite{GANori,braided}, the emitters become decoherence-free in this case yet can still interact with each other in the absence of the direct coupling. However, as show in Fig.~\ref{figb2}(a), the excitation transfer becomes reciprocal in this case, although the coupling phase $\theta=\pi/2$ is nontrivial (we have also checked the dynamic evolutions with other values of $\theta$, which always show reciprocal transfer). Moreover, we can find that in the presence of considerable time delay, the excitation transferred between two emitters still shows a damped oscillation although the damping is much weaker than that in Fig.~\ref{fig2}(b). This is because the emitters are not exactly decoherence-free due to the retarded self interference effects. As $\eta$ (i.e., $d$) decreases gradually, the non-Markovian retarded effect becomes more and more negligible such that the emitters tends to be completely decoherence-free, as shown in Fig.~\ref{figb2}(b).

The results above can be understood as follows. In the Markovian limit ($d/v_{g}\rightarrow0$), the waveguide induced indirect coupling between the two braided giant atoms reads $g_{\textrm{eff}}=-i\frac{\gamma}{2}(3e^{i\phi}+e^{3i\phi})$, which is purely real in the case of $\phi=(m+1/2)\pi$. On the other hand, the direct coupling coefficients $|J|e^{\pm i\theta}$ possess identical real and opposite (vanishing) imaginary parts as long as $\theta\neq m\pi$ ($\theta=m\pi$). In view of this, the overall coupling between $a$ and $b$ is always reciprocal due to the identical strength (modulus) for both directions. For the non-Markovian case here, this can be seen from the Laplace transformation of Eq.~(\ref{eqb1}), i.e.,
\begin{equation}
\begin{split}
s\tilde{c_{a}}(s)-c_{a}(0)=&-\gamma(1+e^{2\varphi})\tilde{c_{a}}(s)-i|J|e^{i\theta}\tilde{c_{b}}(s)\\
&-\frac{\gamma}{2}(3e^{\varphi}+e^{3\varphi})\tilde{c_{b}}(s),\\
s\tilde{c_{b}}(s)-c_{b}(0)=&-\gamma(1+e^{2\varphi})\tilde{c_{b}}(s)-i|J|e^{-i\theta}\tilde{c_{a}}(s)\\
&-\frac{\gamma}{2}(3e^{\varphi}+e^{3\varphi})\tilde{c_{a}}(s),
\end{split}
\label{eqb2}
\end{equation}
where $\tilde{c_{j}}(s)$ $(j=a,\,b)$ is the Laplace transformation of $c_{j}(t)$ and $\varphi=i\phi-sd/v_{g}$. Eq.~(\ref{eqb2}) shows a pair of complex conjugate overall coupling coefficients in the $s$-domain, implying that the excitation transfer should be reciprocal in the absence of other mechanisms that may induce nonreciprocity (such as optical Sagnac effects~\cite{sagnac}). We point out that the nonreciprocal transfer can never be achieved as long as the emitters are ``decoherence-free'' [i.e., the phase accumulation between the two coupling ports of each emitter is $(2m+1)\pi$], even if the separation distances between adjacent coupling ports are not equal or, the coupling ports are not arranged in the braided manner (this can be verified with some algebraic calculations). This is because the waveguide induced indirect coupling is either vanishing or purely coherent (the coefficient is real valued) in this case while the nontrivial coupling phase $\theta$ can only introduce opposite imaginary coupling terms.

\end{document}